%% file: main.tex
\RequirePackage{fix-cm}
\documentclass[smallextended]{svjour3}       % onecolumn (second format)
\smartqed  % flush right qed marks, e.g. at end of proof
\usepackage{graphicx}
\usepackage{chronosys}
\usepackage[multiple]{footmisc}
\usepackage{textalpha} % <--- Greek letters in text
\usepackage{xurl} % load either 'url' or 'xurl' package
\usepackage[numbers]{natbib}
\usepackage{booktabs} 
\usepackage[utf8]{inputenc}
\usepackage{subcaption}
\usepackage{graphicx}
\usepackage{textcomp}
\usepackage{xcolor}
\usepackage{xspace}
\usepackage{listings}
\usepackage{color}

\def\code#1{\texttt{#1}}
\usepackage{multirow}
\usepackage{amsmath}
\usepackage{mathtools}
\usepackage{graphicx,wrapfig,lipsum}
\usepackage{color, colortbl}
\definecolor{LightCyan}{rgb}{0.88,1,1}
\usepackage{threeparttable}
\usepackage{xcolor}    % loads also »colortbl«
\usepackage{tcolorbox}
\usepackage{soul}
\usepackage{relsize}
\usepackage{tikz}
\usepackage{todonotes}
\usepackage[T1]{fontenc}

\usepackage[altpo]{backnaur}
\usepackage{booktabs} % For formal tables
\usepackage{flushend}
\usepackage[normalem]{ulem}
\usepackage{footnote}
\renewcommand{\thefootnote}{\arabic{footnote}}
\usepackage[misc]{ifsym}

\usepackage{paralist}
\definecolor{block-gray}{gray}{0.95}

\newtcolorbox{rqbox}[1][]{
    colback=gray!10,
    colframe=gray,
    arc=1mm,
    boxrule=0.5pt,
    coltitle=black,
    fonttitle=\bfseries,
    title=#1
}

\newtcolorbox{RQ_answer}[2][]
{
	size= title,
	colframe = black,
	colback  = #2!10,
	#1,
}

\newtcolorbox{zitat}[2][]{%
    colback=block-gray,
    grow to right by=-3mm,
    grow to left by=-3mm, 
    boxrule=0pt,
    boxsep=0pt,
    breakable,
    enhanced jigsaw,
    borderline west={4pt}{0pt}{gray},
    title={#2\par},
    colbacktitle={block-gray},
    coltitle={black},
    fonttitle={\large\bfseries},
    attach title to upper={},
    #1,
}

\definecolor{dkgreen}{rgb}{0,0.6,0}
\definecolor{gray}{rgb}{0.5,0.5,0.5}
\definecolor{mauve}{rgb}{0.58,0,0.82}
\input{solidity-highlighting.tex}
\newcommand*{\affaddr}[1]{#1} 
\newcommand*{\affmark}[1][*]{\textsuperscript{#1}}

\newcommand{{\dataset}}{{\sc SolBench}}

\begin{document}

\title{Charting The Evolution of Solidity Error Handling}

%\titlerunning{Short form of title}        % if too long for running head

\author{Charalambos Mitropoulos$^*$ \protect\affmark[1] \and Maria Kechagia$^*$ \and \protect\affmark[2] \and Chrysostomos Maschas \protect\affmark[3] \and Sotiris Ioannidis \protect\affmark[1] \and Federica Sarro \protect\affmark[2] \and Dimitris Mitropoulos \protect\affmark[4] }

%\authorrunning{Short form of author list} % if too long for running head
\def\thefootnote{*}\footnotetext{The first two authors contributed equally to this work.}\def\thefootnote{\arabic{footnote}}

\institute{\Letter $\;\;$  Charalambos Mitropoulos \\   \email{cmitropoulos@tuc.gr}\\\\
 Maria Kechagia \\ \email{m.kechagia@ucl.ac.uk}\\\\
 Chrysostomos Maschas \\ \email{chrysom@noc.grnet.gr}\\\\
 Sotiris Ioannidis \\ \email{sioannidis@tuc.gr}\\\\
 Federica Sarro \\ \email{f.sarro@ucl.ac.uk}\\\\
 Dimitris Mitropoulos \\ \email{dimitro@ba.uoa.gr}\\\\
 \affaddr{\affmark[1] Technical University of Crete} \\
 \affaddr{\affmark[2] University College London} \\
 \affaddr{\affmark[3] National Infrastructures for Research and
  Technology (GRNET)} \\
   \affaddr{\affmark[4] University of Athens}}

\date{Received: date / Accepted: date}
% The correct dates will be entered by the editor

\maketitle

\begin{abstract}
The usage of error handling in Solidity smart contracts
is vital because smart contracts perform transactions
that should be verified.
Transactions that are not carefully handled,
may lead to program crashes and vulnerabilities,
implying financial loss and legal consequences.
While Solidity designers attempt to constantly update
the language with new features,
including error-handling ({\sc EH}) features,
it is necessary for developers to promptly absorb
how to use them.

We conduct a large-scale empirical study on 283K unique
open-source smart contracts
to identify patterns regarding the usage of
Solidity {\sc EH} features over time.
Overall,
the usage of most {\sc EH} features is limited. 
However,
we observe an upward trend ($>$ 60\%) in
the usage of a Solidity-tailored {\sc EH} feature,
i.e.,
{\tt require}.
This indicates that designers of modern programming
languages may consider making
error handling more tailored to the purposes of
each language.
Our analysis on 102 versions of the Solidity documentation
%as well as the analysis of
%smart contracts with
%missing error handling found in our dataset,
indicates the volatile nature of Solidity,
as the language changes frequently,
i.e.,
there are changes on {\sc EH} features once or twice a year.
Such frequent releases may
confuse smart contract developers,
discouraging them to carefully read the Solidity documentation,
and correctly adopt {\sc EH} features.
%Thus, Solidity should consider developer
%tolerance towards the frequency of
%Solidity’s changes.
%e.g., shielding external calls for safe
%transactions between Solidity smart contracts.
Furthermore,
our findings reveal that nearly 70\% of the
examined smart contracts are
exposed to potential failures due to missing error handing,
e.g.,
unchecked external calls.
Therefore,
the use of {\sc EH} features should be further supported
via a more informative documentation containing
(1) representative and meaningful examples
and (2) details about
the impact of potential {\sc EH} misuses.
%or ad-hoc tools that can suggest error-handling usage
%to engage developers.
\end{abstract}

\section{Introduction}
\input{sections/introduction}

\section{Background}
\input{sections/background}
\label{background}

\section{Methodology}
\input{sections/setup}
% \vspace{-2.85mm}
\section{Results}
\label{sec:res}
\input{sections/results}

\section{Threats to Validity}
\input{sections/threats}

\section{Related Work}
\input{sections/rw}

\section{Conclusions}
\input{sections/conclusions}

\begin{comment}
\section{Data Availability}
One can reproduce the results
of our study by accessing
\dataset{} online.\footnote{\url{https://github.com/Solidity-ErrorHandling-Anonymous/solbench}}
We plan to make a corresponding eponymous repository
publicly available upon acceptance.
\end{comment}

%\begin{acknowledgements}
%Maria Kechagia and Federica Sarro are
%supported by the ERC grant no. 741278 (EPIC).  
%\end{acknowledgements}

%\begin{acknowledgements}
%If you'd like to thank anyone, place your comments here
%and remove the percent signs.
%\end{acknowledgements}

% BibTeX users please use one of
\bibliographystyle{spbasic}      % basic style, author-year citations
% \bibliographystyle{spmpsci}      % mathematics and physical sciences
%\bibliographystyle{spphys}       % APS-like style for physics
%\bibliography{}   % name your BibTeX data base
\bibliography{main}

\end{document}

%% file: solidity-highlighting.tex
\usepackage{listings, xcolor}

\definecolor{verylightgray}{rgb}{.97,.97,.97}

\lstdefinelanguage{Solidity}{
	keywords=[1]{anonymous, assembly, assert, balance, break, call, callcode, case, catch, class, constant, continue, constructor, contract, debugger, default, delegatecall, delete, do, else, emit, event, experimental, export, external, false, finally, for, function, gas, if, implements, import, in, indexed, instanceof, interface, internal, is, length, library, log0, log1, log2, log3, log4, memory, modifier, new, payable, pragma, private, protected, public, pure, push, require, return, returns, revert, selfdestruct, send, solidity, storage, struct, suicide, super, switch, then, this, throw, transfer, true, try, typeof, using, value, view, while, with, addmod, ecrecover, keccak256, mulmod, ripemd160, sha256, sha3}, % generic keywords including crypto operations
	keywordstyle=[1]\color{blue}\bfseries,
	keywords=[2]{address, bool, byte, bytes, bytes1, bytes2, bytes3, bytes4, bytes5, bytes6, bytes7, bytes8, bytes9, bytes10, bytes11, bytes12, bytes13, bytes14, bytes15, bytes16, bytes17, bytes18, bytes19, bytes20, bytes21, bytes22, bytes23, bytes24, bytes25, bytes26, bytes27, bytes28, bytes29, bytes30, bytes31, bytes32, enum, int, int8, int16, int24, int32, int40, int48, int56, int64, int72, int80, int88, int96, int104, int112, int120, int128, int136, int144, int152, int160, int168, int176, int184, int192, int200, int208, int216, int224, int232, int240, int248, int256, mapping, string, uint, uint8, uint16, uint24, uint32, uint40, uint48, uint56, uint64, uint72, uint80, uint88, uint96, uint104, uint112, uint120, uint128, uint136, uint144, uint152, uint160, uint168, uint176, uint184, uint192, uint200, uint208, uint216, uint224, uint232, uint240, uint248, uint256, var, void, ether, finney, szabo, wei, days, hours, minutes, seconds, weeks, years},	% types; money and time units
	keywordstyle=[2]\color{teal}\bfseries,
	keywords=[3]{block, blockhash, coinbase, difficulty, gaslimit, number, timestamp, msg, data, gas, sender, sig, value, now, tx, gasprice, origin},	% environment variables
	keywordstyle=[3]\color{violet}\bfseries,
	identifierstyle=\color{black},
	sensitive=true,
	comment=[l]{//},
	morecomment=[s]{/*}{*/},
	commentstyle=\color{gray}\ttfamily,
	stringstyle=\color{red}\ttfamily,
	morestring=[b]',
	morestring=[b]"
}

%% file: sections/introduction.tex
\label{sec:intro}
{\it Smart contracts}~\cite{AH19, MCFLD21, CWAWL22}
are computer programs
stored on a
{\it blockchain}
(i.e., a system
maintaining a record of {\it transactions}
across computers linked in a peer-to-peer
network~\cite{GKWGERC16, KMMMMB18})
that can be used for automating
the execution of transactions
between different parties.
These transactions include
triggering a payment or a service delivery,
registering a vehicle, or issuing a ticket. 

{\it Solidity}~\cite{sol} is a recent object-oriented programming
language, released in 2014,
for developing smart
contracts that run on blockchain platforms such as {\it Ethereum}~\cite{ethereum}.
Ethereum is one of the largest and most popular
decentralised platforms where 1M transactions take place
everyday~\cite{etherium-transactions, econ}.

As other programming languages provide some sort of
{\it error handling} to manage unexpected
errors that may manifest at run-time~\cite{Kin06, WN08, SGH10, ZLH23}, Solidity, also, provides a number of {\it error-handling} ({\sc EH})
{\it features}~\cite{controlstr}, to handle transactions
that can be exposed to potential run-time errors.
%that can lead to failures. 
Such features include standards like {\tt try}--{\tt catch} and {\tt assert},
as well as features developed specifically for Solidity, i.e., {\tt require} and {\tt revert}.

Despite of the provenance of {\sc EH} features in Solidity,
it seems that developers do not fully understand those features and often neglect their usage.
The Nomad bridge attack~\cite{nomad}, which costed millions of dollars,
is just one of the most recent examples of how a correct usage of
{\sc EH} features can prevent critical losses.
In fact,
this attack was due to a missing check for a zero {\tt address}
({\tt 0x00}).
Without this check the contract would mark a zero
{\tt address} as a valid {\tt address} for incoming messages.
Thus,
attackers were able to perform malicious,
yet valid transactions by using the
{\tt 0x00} {\tt address} as a stepping stone.
%\hl{Neglecting to check for a zero {\tt address}
%could enable an attacker
%to send a transaction to an invalid recipient.
%Failing to handle this scenario properly within the smart contract,
%could result in fund theft or other significant damage by an attacker.}
Such an attack could have been avoided if the developers
had used the {\tt require} {\sc EH} feature
provided by Solidity (as we will explain later on,
{\tt require} can be used to verify external inputs before execution).
Amann et al.~\cite{ANN18,SNN19} consider the coding situation where
error handling is ignored (similar to the Nomad attack),
as an {\sc EH} {\it misuse},
i.e., a violation of the specification usage of a programming language's element.
Therefore, due to the serious financial and legal implications
of transaction failures in smart contracts,
the reliable and secure execution of a smart contract
through the {\it correct} usage of {\sc EH} features
should be a top priority for developers of Solidity smart contracts.

Even though there are several studies
that examine the {\sc EH} usage
and its impact for different programming
languages~\cite{BG19,BC15,SGH10,NHT16,KGM16,PZ21},
this is not the case for Solidity.
Given that Solidity is a relative new programming language,
its characteristics and features evolve on a daily basis.
Thus,
focusing on the evolution of its {\sc EH}
features and corresponding documentation would
provide important feedback both for the
designers of Solidity and developers of smart contracts.
While there are several empirical
studies related to Solidity,
focusing, for instance,
on the use of inline assembly~\cite{CGL22},
the performance of program-analysis tools~\cite{DFA20},
and code reuse ~\cite{CLZ21},
to the best of our knowledge,
there is no empirical study that examines
the evolution of Solidity EH features.
Our work aims to fill this research gap.

In this paper, we present the first large-scale
empirical study on 283K unique open-source
Solidity smart contracts
to understand how developers use Solidity {\sc EH} features over time
and make suggestions to facilitate the features' correct use.
Specifically,
we aim to answer the following research questions.

{\bf RQ1: What is the usage frequency of Solidity error-handling features?}
Our work considers the usage of different
Solidity {\sc EH} features and categorises them.
Answering RQ1 will show us the extent to which
developers use error handling in smart contracts that are deployed in production.
Such information can highlight, for instance,
{\sc EH} features that are scarcely
used by developers, and raise awareness for
improving the specification usage of these features in the Solidity documentation.

{\bf RQ2:
How does the usage of error-handling features evolve
as new Solidity versions are released?}
A closer look at the different Solidity
releases will shed light on the major changes regarding
{\sc EH} features and their characteristics.
Answering RQ2 will help us observe new or
deprecated {\sc EH} features
and form a
timeline that can reveal the landscape of
Solidity's evolution.

{\bf RQ3:
How often do developers use Solidity error-handling features,
in their smart contracts, over time?}
We examine how often developers use Solidity {\sc EH} features over time,
and as new versions of Solidity are released.
Answering RQ3 will help us identify potential trends and observe
usages of error handling before and after the
introduction of important modules.
Then,
we will be able to understand whether developers
follow the upgrades introduced in every new Solidity
version and how these changes impact the usage
of {\sc EH} features in Solidity smart contracts.

{\bf RQ4:
Which are the types of error-handling misuses in Solidity
and what is the frequency of each type?}
Our goal here is two-fold.
First, we want to identify different categories of {\sc EH} misuses and their potential impact.
Consider that developers can read about error handling and how to use it in the Solidity documentation. However,
the existing Solidity documentation does not include examples illustrating the different coding situations where {\sc EH} features should be used.
This means that Solidity documentation does not clarify the impact in case of a misuse.
Second, we intend to examine how often developers fail to use {\sc EH} features when they should.
Our findings can be used to make suggestions
that engage developers in {\sc EH} usage.

{\bf RQ5:
How do error-handling misuses evolve in the Solidity context?}
We want to examine whether the number of {\sc EH} misuses changes over time. 
Answering RQ5 will help us understand potential trends (e.g., a sudden increase) and correlate them with the corresponding Solidity releases and their features.

To answer our research questions,
we have developed {\sc SolBench},
an approach that enables the automated collection and
analysis of real-world smart contracts.
Using {\sc SolBench} one can
(1) automatically collect publicly available
smart contracts on a regular basis;
(2) measure the evolution of
{\sc EH} feature usages;
(3) observe important
changes in Solidity error handling;
and (4) identify
{\sc EH} misuses and their frequencies.
We carry out our empirical study on a dataset consisting of 1.7M source
files and 283K of unique smart contracts
(some of which were made available in previous work ~\cite{CGL22}).

Overall, our findings show that the most used {\sc EH}
feature is {\tt require} (83.15\%),
while the least used {\sc EH} feature is {\tt assert} (3.82\%) ({\bf RQ1}).
Significant changes in Solidity {\sc EH} features
include the deprecation of {\tt throw} early on,
and the introduction of {\tt require} and {\tt try}--{\tt catch}
at different points in time ({\bf RQ2}).
Furthermore, {\tt require} has the highest usage increase across
Solidity versions and over the years.
When {\tt try}--{\tt catch} was introduced in Solidity,
the number of its usage increased instantly, becoming equal
to those of {\tt revert} and {\tt assert} ({\bf RQ3}).
Popular misuses over time involve missing {\sc EH}
features to check if either an {\tt address} type
(i.e., the {\tt address} of a block) is valid,
or the call performed to an external contract was successful ({\bf RQ4}).
We also observe a positive increase over time only in the adoption
of some specific {\sc EH} features ({\bf RQ5}).
Notably, an upward trend ($>$ 60\%) in the usage of
Solidity-tailored {\sc EH} features (i.e., {\tt require})
suggests that designers of modern programming languages may
want to consider devising {\sc EH} features that are more
specific to each language's purposes.
%e.g., shielding external calls for safe transactions between Solidity smart contracts.

Our work makes the following contributions:
\begin{itemize}
\item {\bf Approach.}
Our methods and dataset
are publicly available~\cite{solbench},
%\footnote{\url{https://github.com/Solidity-ErrorHandling-Anonymous/solbench}}
and can be used to replicate our study
and for further research in the field.

\item {\bf Study.}
We provide the first empirical study,
to the best of our knowledge,
on the evolution of error handling in Solidity,
by analysing real-world Solidity smart contracts.
\item {\bf Misuses.}
We identify seven categories of misuses in Solidity error handling,
provide representative examples, and discuss their implications for users.
Based on our findings, we make suggestions for the improvement of Solidity and its documentation.
\end{itemize}

{\bf Data availability.} One can reproduce the results
of our study by accessing
\dataset{}
online~\cite{solbench}.
We plan to make a corresponding eponymous repository
publicly available upon acceptance.
%\footnote{\url{https://github.com/Solidity-ErrorHandling-Anonymous/solbench}}

%% file: sections/background.tex
\label{sec:background}

{\bf Error Handling in Solidity}.
Solidity is based on~\code{solc},
the {\it standard} Solidity compiler,
which counts
$\sim$100 releases since 2015~\cite{sol}.
{\tt solc} offers several mechanisms
%to prevent transaction failures
including
{\sc EH} features and
%\code{solc} offers various components
%to support smart contract functionalities
%including
%such as {\it inline assembly},
%an {\it Application Binary Interface} ({\sc abi})
%that facilitates the communication with
%smart contract bytecode,
%and
the {\sc SMTChecker},
a built-in
{\it formal verification module}~\cite{smt}.
%In this study,
%we focus on {\sc EH} features.

%\code{solc} supports a number of
%{\sc EH} mechanisms to prevent
%transaction failures.
%Solidity also includes exception-handling mechanisms
%({\it error handling})
%to prevent transaction failures.
According to Solidity's API reference
documentation (hereafter, {\it Solidity documentation})~\cite{controlstr},\footnote{We refer to Solidity versions~$\leq$ \code{0.8.19} since \code{0.8.19} is the latest version found available at the time of our study.}
Solidity uses state-reverting exceptions
to handle potential runtime errors.
When an exception manifests in a sub-call,
it is automatically propagated unless it is caught
via error handling.
Solidity offers the {\sc EH} features
presented in the following paragraphs.

Two {\sc EH} features
tailored to Solidity are
{\tt require} and {\tt revert}.
{\tt require} checks
for programming conditions,
and throws an exception
when particular
conditions are not met.
%In the Solidity context,
%this feature can be used to
%``demand'' something before
%availing the service to a user.
%{\tt revert} is used in a similar fashion.
%Nevertheless, it does not evaluate statements and does not %depend on any state or other statements.
{\tt revert} has the same semantics as
the {\tt throw} keyword used in older Solidity versions
(it was removed in version 0.5.0 as we will see later
in the paper),
and in other programming languages such as Java~\cite{javathrow}.
If {\tt revert} is triggered,
an exception is thrown,
along with the return of
{\it gas} (i.e., the fee required to
perform a transaction on the Ethereum blockchain),
and reverts to its original state.

\code{solc} also supports two
well-known {\sc EH} features
inspired by other programming languages
such as Java,
namely: {\tt try}--{\tt catch}
and {\tt assert}.
A {\tt try} statement allows
developers to define a block of
Solidity code
to be tested for errors,
while it is being executed.
{\tt catch}
allows the definition of a block of code to be executed
if the error that occurred was in the
corresponding {\tt try} block.
{\tt assert}
can be employed to check for specific conditions
and if the conditions are not met it throws an exception.

\noindent
{\bf Error-Handling (EH) Misuses}.
The usage specification
of the Solidity {\sc EH} features,
i.e.,
when and how {\sc EH} features should be used,
is defined in the Solidity documentation.
An {\sc EH} misuse occurs
when the developers of smart contracts violate
the usage specification of the {\sc EH} features.
Such misuses may affect the reliable execution
and the security of a smart contract.
Specifically,
if developers ignore the Solidity
{\sc EH} features,
particular exceptions can manifest
causing detrimental effects,
such as the unexpected termination of a smart contract.
{\sc EH} misuses may also introduce security vulnerabilities~\cite{CPNX20, SWC-104, SWC-101, RP15}
that can lead even to a DoS (Denial of Service) attack~\cite{SWC-113}.
For instance,
recall the Nomad Bridge attack~\cite{nomad}
(discussed in Section~\ref{sec:intro})
that led to a \$190M loss.
This attack was based on a missing {\tt address}-zero check,
which should have been performed via error handling.
%Therefore,
%it is essential for developers to fix
%{\sc EH} misuses in their smart contacts before deployment.

According to a recent
taxonomy of API
misuses~\cite{ANN18},\footnote{An Application Programming Interface ({\sc api})
can be considered as
a publicly available bundle of interfaces,
classes, and methods,
that client programs can call or implement.
An API {\it usage} refers to any call of one
or more methods of either old or new versions of an
API, i.e., $API=[m1,...,mk]$~\cite{FXO19}.
An API misuse is a violation of the usage
specification of that API.}
there are two categories referring to {\sc EH} misuses:
(1) missing usage of error handling
and (2) redundant usage of error handling.
In this study, we examine coding situations
where error handling is missing (hereafter, {\it {\sc EH} misuses}).
Notably,
the aforementioned work indicates that
the absence of {\sc EH} usage
can have a greater negative impact on the
functionality of a program
than the redundant {\sc EH} usage.
Even though the work of Amman et al.~\cite{ANN18}
refers to API-misuse categories for Java APIs,
we argue that the identified categories
are generic enough to be applied to
APIs for different programming languages,
including Solidity.
%that share similar behaviours to Java.

Listing~\ref{misuse} presents a code excerpt,
where {\sc EH} is missing,
i.e., the developer should have used {\tt require}
to check that the {\tt account} (see the first argument in line 2)
is not a zero {\tt address}~\cite{controlstr}.
Recall that a similar check was missing
in the Nomad bridge attack~\cite{nomad}.
\vspace{-3mm}
\begin{lstlisting}[float, basicstyle=\ttfamily, caption=Missing zero-{\tt address} check via {\tt require}., label=misuse, belowskip=-0.1\baselineskip]{1mm}
//contract ScalpexToken
function _burn(address account, uint256 amount) internal virtual {
    _beforeTokenTransfer(account, address(0), amount);
    _balances[account] = _balances[account].sub(amount, "ERC20:burn amount exceeds balance");
    _totalSupply = _totalSupply.sub(amount);
    emit Transfer(account, address(0), amount);
}
\end{lstlisting}
%\vspace{-4mm}

%% file: sections/setup.tex
%To examine the evolution of
%error handling in the context of Solidity,
%we develop an approach which we call
%{\sc SolBench}.
%Using {\sc SolBench} we are able to
%collect real-world smart contracts incrementally,
%and analyse them in a well-defined manner.
%In the following sections,
%we present the key steps and aspects of our approach.
%highlighting its key aspects.

%In the following sections,
%we present the key steps and aspects of our approach,
%{\sc SolBench}.

\subsection{{\sc SolBench}}
\label{sec:setup}

\begin{figure}[t]
    \centering
	\includegraphics[scale=0.42]{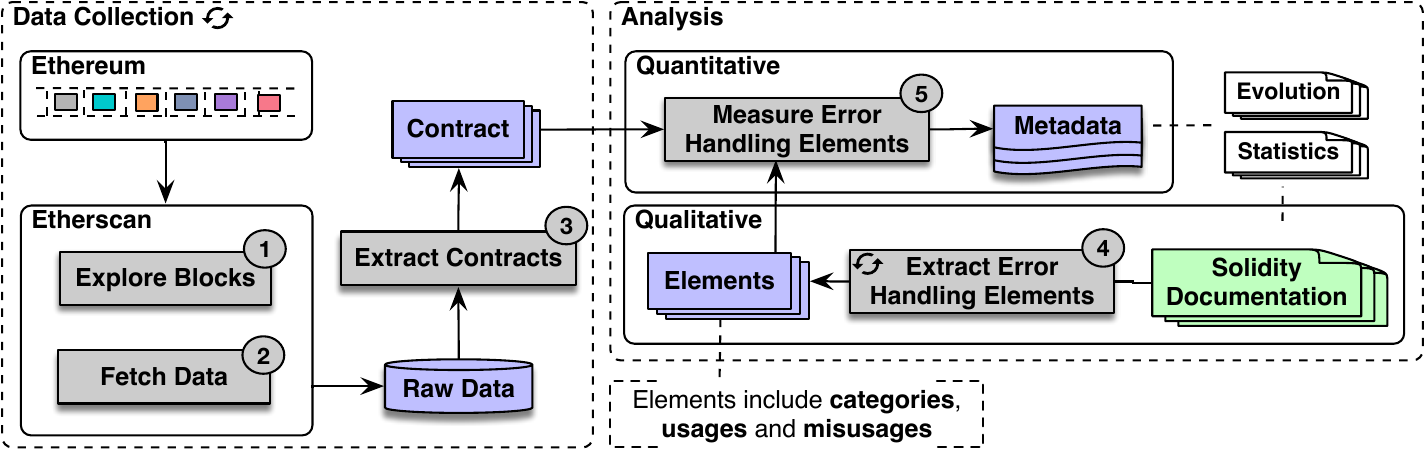}
	%\vspace{-1.5mm}
    \caption{Overview of {\sc SolBench}.}
    \label{fig:overview}
    %\vspace{-1.5mm}
\end{figure}

Figure~\ref{fig:overview} illustrates
the overview of \dataset{}.
First,
we utilise {\it Etherscan} ~\cite{etherscan},
a well-established block explorer for Ethereum.
Specifically,
we use its API to examine blocks
and identify smart contract addresses for a given time period
(\textcircled{1}).
We store the addresses and then employ again the Etherscan API
to download the corresponding smart contracts (\textcircled{2}).
Since not all developers make their smart contract source code publicly available,
we only retain open-source smart contracts for our analysis.
Further,
when a smart contract is downloaded,
it is accompanied by various metadata and
additional files (including dates,
information about the authors and more).
We discard such files, irrelevant to our study,
and store only the smart contract's source code
in a dataset, as we describe in Section \ref{sec:dataset} (\textcircled{3}).
Observe that these three steps happen in an iterative fashion.
To do so,
we manage to automatically
fetch smart contracts every day
and expand our dataset incrementally.

We thoroughly study the documentation of
the various Solidity releases to extract
information about error handling and its use
(\textcircled{4}).
We describe the protocol used to this end
in Section~\ref{sec:protocol}.
Finally,
we conduct static analysis on the smart contract's
source code to identify {\sc EH} uses,
{\sc EH} misuses and their evolution (\textcircled{5}),
based on a set of heuristic rules described in Section~\ref{sec:rules}.

\subsection{Dataset}
\label{sec:dataset}

\begin{table}[t]
\centering
\caption{Descriptive statistics of our dataset.}
\vspace{-1mm}
\label{table:stats}
\resizebox{0.97\linewidth}{!}{
\begin{tabular}{lr}
\toprule
Start Block & 47,205 (7th of August, 2015)\\
End Block & 16,993,877 (8th of April, 2023) \\
\hline
Total Source files & 2M  \\
Total Unique Source files & 1.7M\\
 \hline
 Total Number of Functions& 168M\\
 Total Number of Modifiers&4.03M\\
 Total Number of Constructors&672K\\
 \hline
 {\bf Total Unique Smart Contracts} & {\bf 283K} \\
 \hline
 Contracts with at least one EH feature & 215.8K \\
 \hline
 Average LOC per \textit{.sol} file & 105.1 \\
 LOC & 51.2M\\
\bottomrule
\end{tabular}}
\end{table}

We curate a dataset containing 283K unique smart contracts.
The smart contracts use \code{solc} versions from \code{v0.1.2} to
\code{v0.8.19}.\footnote{A range of \code{solc} versions that
are compatible with a smart contract are defined in the first
lines of the contract.}
%\hl{Notably, 
%the identification of these Solidity versions 
%was accomplished during the parsing of the smart contracts within our dataset,
%as these version details
%are embedded in the opening lines of the source code.}
We consider Solidity versions from \code{v0.1.2} and on-wards,
because previous versions did not have an accompanying documentation
to examine.
\code{v0.8.19} was the latest found available at the time of our study.
Apart from the smart contracts
we collect
(in the way discussed in Section~\ref{sec:setup}),
our dataset incorporates another well-established benchmark
used in previous work that examines the usage of inline assembly in Solidity~\cite{CGL22}.

During our contract collection,
we perform corresponding checks for potential duplicates
which we filter out.
Up until now,
we have identified
numerous duplicates,
i.e.
$\approx300$K files.
%While performing a number of
%integrity checks on the smart contracts
%we gathered and the contracts contained in
%the existing benchmark used~\cite{CGL22},
%we observed the there were numerous duplicates,
%which we filtered out ($\approx300.000$ files).
Examples of duplicate code include common third-party APIs
such as {\tt ERC721}~\cite{ERC721}
and {\tt SafeERC20}~\cite{SafeERC720}.
Also, in many occasions,
developers deploy their
smart contracts together with other existing smart contracts,
hence, the duplicates.
Further, the benchmark by~\citet{CGL22}
also used in our benchmark,
incorporates empty smart contracts
(without source code).
Such cases would not provide meaningful
information in our case, and, thus, we exclude them from our work.

Table~\ref{table:stats}
displays descriptive statistics for our dataset,
covering contracts from August 2015 to April 2023.
The dataset comprises 1.7 million unique source files.
It is worth noting that a single smart contract may encompass multiple source files with a \textit{.sol} extension.
On average,
a contract consists of around 7.25 \textit{.sol} files, 
with a median value of 6 files.
Additionally,
the average lines of code per \textit{.sol} file is approximately 105.1.
In total,
the dataset contains approximately 283,000 unique smart contracts
and 76\% of these contracts include at least one {\sc EH} feature.

\subsection{Documentation Analysis}
\label{sec:protocol}

An important part of our approach
involves the analysis of the Solidity documentation.
To this end,
two authors independently examined:
\textcircled{a} the documentation of the
latest Solidity version at the time of our study
(i.e., \code{v0.8.19}) to identify
elements such as usage categories
for the {\sc EH} features,
and
\textcircled{b} the documentation of each \code{solc} version
from \code{v0.1.2} to \code{v0.8.19}
to record changes related to {\sc EH}
usages.
Consider that \code{v0.1.2} is the first \code{solc}
release that includes meaningful documentation files
(i.e., previous versions incorporated insufficient material,
which would not lead to consistent results).
\textcircled{a} was straightforward
since the documentation is clear regarding the usages of
all {\sc EH} features except for {\tt assert},
where we form a number of new categories.
We further explain these new categories in
Section~\ref{sec:categories}.
Through \textcircled{a}, we are able to answer
RQ1 and RQ3 to RQ5.
Furthermore,
\textcircled{b} helps us to answer RQ2, but also
contributes to the results of other RQs
as described in Section~\ref{sec:res}.

The authors discussed their
observations and findings,
until they reached consensus.
The procedure was repeated three times.
During these iterations,
the authors revisited,
and updated their observations.
Finally,
two additional authors verified the resulting findings
and discussed potential conflicts with the
initial authors until agreement.
Note that having this additional
step was critical in cases where
the two authors that conducted the manual analysis could not
reach consensus.

\subsection{Error-Handling Usage Categories}
\label{sec:categories}

Studying the Solidity documentation
(\code{v0.8.19})
we identify the following {\sc EH}
usage categories.

{\tt require} can be used to evaluate
(1) function arguments 
and (2) external calls that are used to call
functions from other contracts.
Consider the code excerpt in Listing~\ref{extcalls}. 
%\vspace{-2mm}
\begin{lstlisting}[float, basicstyle=\ttfamily, caption=Use of {\tt require} in external call., label=extcalls, belowskip=-0.8\baselineskip]{3mm}
//contract: TellerV2Context
function setTrustedMarket(uint256 _mId, address _forwarder) external {
    require( marketRegistry.getMarketOwner(_mId) == _Sender(), "Caller must be the marketowner");
}
\end{lstlisting}

% \vspace{3mm}
\noindent
The \code{setTrustedMarket} function
calls the external function \code{getMarketOwner}
from the contract \code{marketRegistry}.
In this case,
\code{require}
is used to test if the call
works as it is supposed to.

{\tt try}--{\tt catch} can be used to:
(1) catch and handle failures from external calls,
in the same way as \code{require} does,
and
(2) evaluate the creation of external contracts.
The last case occurs when a contract creates an external contract
and the developer has to be sure that the creation is done correctly.

{\tt revert} can be called as (1) a function and as
(2) a statement
to throw an error.
%(i.e., it works as the {\tt throw} statement in Java~\cite{javathrow}).
Consider the code excerpt in Listing~\ref{revertfunction}.
%\vspace{-2mm}
\begin{lstlisting}[float, basicstyle=\ttfamily, caption=Use of {\tt revert} with custom error., label=revertfunction, belowskip=-0.8\baselineskip]
//contract: Strategy
error OnlyOwner();
function updateOperatorFilterRegistryAddress(address newRegistry) public virtual { 
    if (msg.sender != owner()) {
        revert OnlyOwner();
    }
}
\end{lstlisting}

\noindent
In that case,
the developer defines its own error function
called \code{OnlyOwner()}.
If the \code{if} statement is {\tt true}, the
\code{revert} function will be called
providing an error message defined by the developer.
In the same manner,
in the case of a \code{revert} statement,
the error message is provided
inside quoted marks:
\code{revert("...")}.
    
{\tt assert} can be used to check specific coding situations
listed in the Solidity documentation~\cite{soldoc}.
Each case has a corresponding
error code.
After analysing all situations,
we grouped them into five categories.
The following list mentions
the codes of each category considered in our study:
(1) arithmetic overflow/underflow
(with error code \code{0x11}),
(2) division by zero (\code{0x12}),
(3) array operations (\code{0x22}, \code{0x31}, \code{0x32}, \code{0x41}),
(4) program logic (\code{0x01}, \code{0x51}), and
(5) {\tt enum} type conversion (\code{0x21}).
An example belonging to
the {\tt enum} type conversion category,
involves the code excerpt in Listing~\ref{enumstype}.
%\vspace{-2mm}
\begin{lstlisting}[float, basicstyle=\ttfamily, caption=Use of the {\tt enum} type conversion., label=enumstype, belowskip=-0.8\baselineskip]{3mm}
//contract: Hakiro
enum Step { Before, PublicSale, WhitelistSale, SoldOut, Reveal}
function setStep(uint _step) external onlyOwner {
    assert(sellingStep = Step(_step));
}
\end{lstlisting}

\noindent
Here, the developer attempts to convert a \code{uint}
value named \code{\_step},
into the corresponding \code{Step} {\tt enum} value.
To ensure that the process is done correctly,
the \code{assert} function is called.
Consider that the \code{0x00} code (i.e., {\it ``generic compiler inserted panics"})
is not assigned to any category,
because it is too generic to classify.

\begin{figure}[t]
\small
\hspace*{-10cm} 
\begin{bnf*}
    \bnfprod{$u \in Usages$}
     {\bnftd{eh(c.f com (t | f), t | er)} \bnfor
     \bnftd{eh(t com t, t) } \bnfor} \\[-0.8mm]
     \bnfmore*{
     \bnftd{eh(t \textit{op} t)} \bnfor
     \bnftd{eh(er | t)} \bnfor
     \bnftd{{eh(arr.l com t) \& u } }\bnfor} \\[-0.5mm]
     \bnfmore*{
     \bnftd{eh(e(t))}\bnfor
     \bnftd{eh(arr.{\bf pop()})} \bnfor
     \bnftd{eh({\bf new} c.f)} \bnfor} \\[-0.5mm]
     \bnfmore*{
     \bnftd{eh(t \textit{com} t)}
     } \\[-0.5mm]
     \bnfprod*{$eh \in Features$}
     {\bnftd{\bf{require}} \bnfor
     \bnftd{\bf{revert}} \bnfor
     \bnftd{\bf{try}} \bnfor
     \bnftd{\bf{assert}}} \\ [0.5mm]
     \bnfprod*{$c \in Smart Contracts$}
     {\bnftd{\text{is the set of available smart contracts}}}\\[-0.5mm]
     \bnfprod*{$f \in Functions$}
     {\bnftd{\text{is the set of functions}}}\\[-0.5mm]
     \bnfprod*{$e \in Enum Type$}
     {\bnftd{set of {\bf enum} types}}\\[-0.5mm]
     \bnfprod*{$er$}
     {\bnftd{\bf{Error}} \bnfor
    \bnftd{\bf{Panic}} \bnfor
    \bnftd{CustomError}
    }\\[-0.5mm]
     \bnfprod{$arr \in arrays$}
    {\bnftd{\bf{storage}} \bnfor
    \bnftd{\bf{memory}}}\\[-0.5mm]
     \bnfprod*{$l$}
    {\bnftd{\text{array length}}}\\[-0.5mm]
     \bnfprod*{$t \in Solidity Types$}
    {\bnftd{\bf{str}} \bnfor
    \bnftd{\bf{uint}} \bnfor
    \bnftd{\bf{int}} \bnfor
    \bnftd{\bf{bool}} \bnfor
    \bnftd{\bf{address}}
    }\\[-0.5mm]
     \bnfprod*{$op$}
    {\bnftd{\bf{+}} \bnfor
    \bnftd{\bf{-}} \bnfor
    \bnftd{\bf{*}} \bnfor
    \bnftd{\bf{/}}
    }\\[-0.5mm]
    \bnfprod*{$com$}
    {\bnftd{\bf{<}} \bnfor
    \bnftd{\bf{<=}} \bnfor
    \bnftd{\bf{>}} \bnfor
    \bnftd{\bf{>=}} \bnfor
    \bnftd{\bf{==}} \bnfor
    \bnftd{\bf{!=}} \bnfor
    \bnftd{\bf{!}}
    }\\[-0.5mm]
\end{bnf*}
% \vspace{-6mm}
\caption{The abstract syntax of the heuristic rules that represent the usages
of the different error-handling (EH) features,
according to the Solidity documentation (v0.8.19).}
\label{fig:syntax}
%\vspace{-0.5mm}
\end{figure}

\subsection{Heuristic Rules}
\label{sec:rules}
To automatically detect
{\sc EH} usages
we define a set of heuristic rules.
The rules are based on the
aforementioned usage categorisation (see Section~\ref{sec:categories})
and the examination of real-world
Solidity smart contracts that
contain {\sc EH} features.
Furthermore,
they are generic and can be easily expanded.
Note that our heuristic rules
can also be employed to identify {\sc EH} misuses because
the denial of a heuristic
indicates a potential misuse.
Finally,
we manually validate the effectiveness
of our heuristics
following the steps we describe
in Section~\ref{sec:threats}.

Figure \ref{fig:syntax},
presents the set of our heuristic rules.
Each element of the {\it Usages} set,
requires an {\sc EH} feature 
from the {\it eh} set: 
$\{require,revert,try,assert\}$.
When we detect one of the {\it Usages} patterns
in a smart contract that exists in our dataset,
it means that we identify a (correct) usage 
of an {\sc EH} feature.
Every rule in {\it Usages}
corresponds to a usage category presented
in Section \ref{sec:categories}.
Consider the rule:
\bnftd{eh(t com t)}.
This can lead to the detection of
two patterns of a program logic check,
namely: {\tt require(a >= b)} | {\tt assert(c == d)}.
Futhremore,
through \bnftd{eh(t \textit{op} t)},
we can detect both (1) overflows / underflows
and (2) division by zero checks
(e.g., {\tt assert(a * b)} | {\tt assert(a / b)}.

As an example,
consider Listing~\ref{extcalls}
where an external call takes place.
In this case,
the heuristic rule
that checks for the correct
usage of the {\tt require} feature
is the following:
\bnftd{{eh(c.f com (t | f), t | er)}}.
Specifically,
the rule searches for a pattern where
the output of an external function \bnftd{c.f}
(\code{getMarkeOwner} in our case)
is compared to
(\bnftd{com})
either the output of another function,
\bnftd{f}
(\code{\_Sender()} in our example),
or another type, \bnftd{t}
(e.g., a message in a string).
After the comparison,
another object (\bnftd{t})
is expected,
i.e. the exception thrown by the feature
(in our case a string).

A misuse appears when a heuristic rule
is {\it not} valid.
Consider a situation where
an external function call
(\bnftd{c.f}) such as
{\tt marketRegistry.getMarketOwner} in Listing~\ref{extcalls},
is detected.
In this case,
if {\tt require} is not used to check the call,
a misuse is identified.
In our supplementary material,
we provide more details regarding the rules we use
to identify misuses.
%and corresponding examples.

%% file: sections/results.tex
%We present the findings of our
%methods aimed at addressing our research questions.
%\vspace{0.7mm}
%In the following sections,
%we present the results of our analysis,
%using {\sc SolBench},
%aimed at addressing our research questions.

%\vspace{1mm}
%\subsection{RQ1: Usage of Solidity Error-Handling Usages}
\subsection{RQ1: Frequencies of Error-Handling Usages}
\label{sec:usageoferrorhandling}
\input{sections/rq1}

\subsection{RQ2: Evolution of Error-Handling Features in the
Solidity Compiler}
\label{sec:compiler_changes}
\input{sections/rq2}

\subsection{RQ3: Evolution of Error-Handling Usage in Smart Contracts}
\label{sec:usage_evol}
\input{sections/rq3}

\subsection{RQ4: Categories of Error-Handling Misuses and Frequencies}
\label{sec:rq3}
\input{sections/rq4}

\subsection{RQ5: Evolution of Error-Handling Misuses in Smart Contracts}
\label{sec:rq5}
\input{sections/rq5}

\subsection{Key Take-aways and Suggestions}
\label{sec:discussion}
\input{sections/discussion}

%% file: sections/rq1.tex
\begin{table}[t]
\centering
\caption{RQ1. Appearances (\#) and frequencies (\%)
of error-handling (EH) usage categories.
%We include both overall and per category results.
}
\vspace{-1mm}
\resizebox{0.99\columnwidth}{!}{
\begin{tabular}{ l | l | r | r | r }
    \toprule
   {\bf EH Features} & {\bf Usage Categories} & {\bf (\#)} & {\bf (\%)} & {\bf Total (\%)} \\
   \midrule
   \multirow{2}{*}{{\tt require}}& function arguments&38,059& 56.35&
   %\multirow{2}{*}{67,600}\\
   \multirow{2}{*}{83.15}\\
    &external calls&29,541&43.75\\
   \hline
   \multirow{2}{*}{{\tt try}--{\tt catch}}&external calls&4,135&70.25
   %&\multirow{2}{*}{5,891}\\
   &\multirow{2}{*}{7.25}\\
    &external contract creation&1,756& 29.83\\
   \hline
   \multirow{2}{*}{{\tt revert}}&functions&3,821& 81.22
   %&\multirow{2}{*}{4,703}\\
   &\multirow{2}{*}{5.79}\\
    & statements&885&18.88\\
   \hline
   \multirow{5}{*}{{\tt assert}}& overflow / underflow &1,217&46.93
   %&\multirow{4}{*}{3,102}\\
   &\multirow{5}{*}{3.82}\\
    & division by zero &984 & 37.94\\
    & array οperations &218&8.43\\
    & program logic &107 &4.12\\
    & {\tt enum} type conversion &67&2.58\\
    \toprule
\end{tabular}
}
\label{table:sub_usages}
\end{table}
%The {\bf goal} of RQ1 is to understand patterns of correct usages of Solidity {\sc EH} features, and identify such correct usages in real-world Solidity smart contracts, such as the ones in our dataset.

Table~\ref{table:sub_usages}, shows for each {\sc EH} feature,
the frequency of each usage category identified.
We observe that the most used {\sc EH} feature is {\tt require} (83.15\%). 
Notably, {\tt require} is introduced specifically for Solidity
and is not supported by other programming languages. 
The remaining {\sc EH} features, {\tt assert},
{\tt revert}, and {\tt try}--{\tt catch}, are less used, i.e., $<$ 10\% per feature.

We also analyse the context that each {\sc EH}
feature is used for (i.e., usage category).
{\tt require} is more used for evaluating function
arguments (56.3\%) than for external calls (43.7\%).
{\tt try}--{\tt catch} is mostly used to handle a potential
failure in an external call (70.2\%) and less used to check
whether the creation of an external contract is successful or not.
Furthermore, {\tt revert} function is utilised more
(81.2\%) than the {\tt revert} statement (18.8\%).
This is meaningful because developers can
write specific messages (e.g., ``Not enough funds")
through {\tt revert} functions,
and facilitate debugging.
%which developers can reuse throughout a contract
%On the contrary,
%if they use {\tt revert} statements
%they must write the same message
%over and over again.
Finally,
{\tt assert} is mostly used to
identify potential
arithmetic overflows and underflows (46.93\%)
and  division by zero (37.94\%).
By contrast,
{\tt assert} is least used to 
check for valid array operations (8.43\%),
examine {\tt enum} type conversions (2.58\%),
or evaluate program-specific conditions (4.12\%).

\begin{rqbox}
	{\bf Answer to RQ1:} The most used {\sc EH} feature is {\tt require} (83.15\%),
    while the least used is {\tt assert} (3.82\%).
    {\tt try}--{\tt catch} is mostly used
    to evaluate external calls ($>$70\%).
    {\tt require} is slightly more used for evaluating function arguments (56.3\%) than for external calls
    (43.7\%).
    {\tt assert} is mostly used
    for checking overflows / underflows ($>$35\%) and
    division by zero ($>$30\%).
    {\tt revert} is mostly used as a function
    ($>$80\%).
\end{rqbox}

%% file: sections/rq2.tex
\label{sec:rq2}

\begin{table}[t]
\centering
\caption{RQ2. Error-handling (EH) changes across different Solidity versions.}
\vspace{-1mm}
\resizebox{0.99\linewidth}{!}{
\begin{tabular}{ r | l | l }
\toprule
   \textbf{Version} & \textbf{Date} & \textbf{EH Change} \\
\midrule
   0.1.3 & Sep 23, 2015 & Introduction of \code{throw} \\
   0.4.0 & Sep 8, 2016 & Specification of the compiler version via \code{pragma} \\
   \multirow{3}{*}{0.4.10} & \multirow{3}{*}{Mar 15, 2017} & Introduction of \code{require(condition)} and \code{assert(condition)} \\
    && Support of \code{revert()} as an \code{OPcode} \\
    && to abort with rolling back, but not consuming all gas. \\
   0.4.13 & Jul 6, 2017 & Deprecation of \code{throw()} due to \code{require()}, \code{assert()}, \code{revert()} \\
   0.4.16 & Aug 24, 2017 & Automated support for checking overflows and \textsc{assert} \\
   0.4.22 & Apr 17, 2018 & Specification of Error in \code{require} and \code{revert} \\
   0.6.0 & Dec 18, 2019 & Introduction of \code{try}--\code{catch} \\
   0.6.9 & Jun 4, 2020 & \code{\textsc{SMTChecker}} supports \code{require} and \code{assert} \\
   0.7.2 & Sep 28, 2020 & \code{\textsc{SMTChecker}} supports \code{revert()} \\
   0.8.0 & Dec 16, 2020 & Introduction of \code{Panic(uint)} and \code{Error(string)} \\
   0.8.1 & Jan 27, 2021 & Catch and decode \code{Panic(uint)} and \code{Error(string)} in \code{try}--\code{catch} \\
   0.8.4 & Apr 21, 2021 & Start deprecating \code{pragma experimental \textsc{SMTChecker}} \\
   \multirow{2}{*}{0.8.7} & \multirow{2}{*}{Aug 11, 2021} & Enabling \code{\textsc{SMTChecker}} to check for overflows / underflows \\
   && Running \code{\textsc{SMTChecker}} by default \\
   \bottomrule
\end{tabular}
}
\label{table:signif_changes}
\end{table}

% \startchronology[startyear=2000,stopyear=2025,dates=false,arrow=false,height=0.5ex]
% \chronoevent[markdepth=1.5cm,color=red]{2001}{0.1}
% \chronoevent[markdepth=2.5cm,color=blue]{2005}{0.2}
% \chronoevent[markdepth=3.5cm,color=red]{2010}{1.0}
% \chronoevent[markdepth=4.5cm,color=blue]{2015}{2.0}
% \chronoevent[markdepth=5.5cm,color=red]{2020}{3.0}
% \stopchronology
\vspace{-2mm}
\begin{figure*}[t]
%\hspace*{-2cm}    
    \centering
	\hbox{\hspace{-1.8em}\includegraphics[scale=0.20]{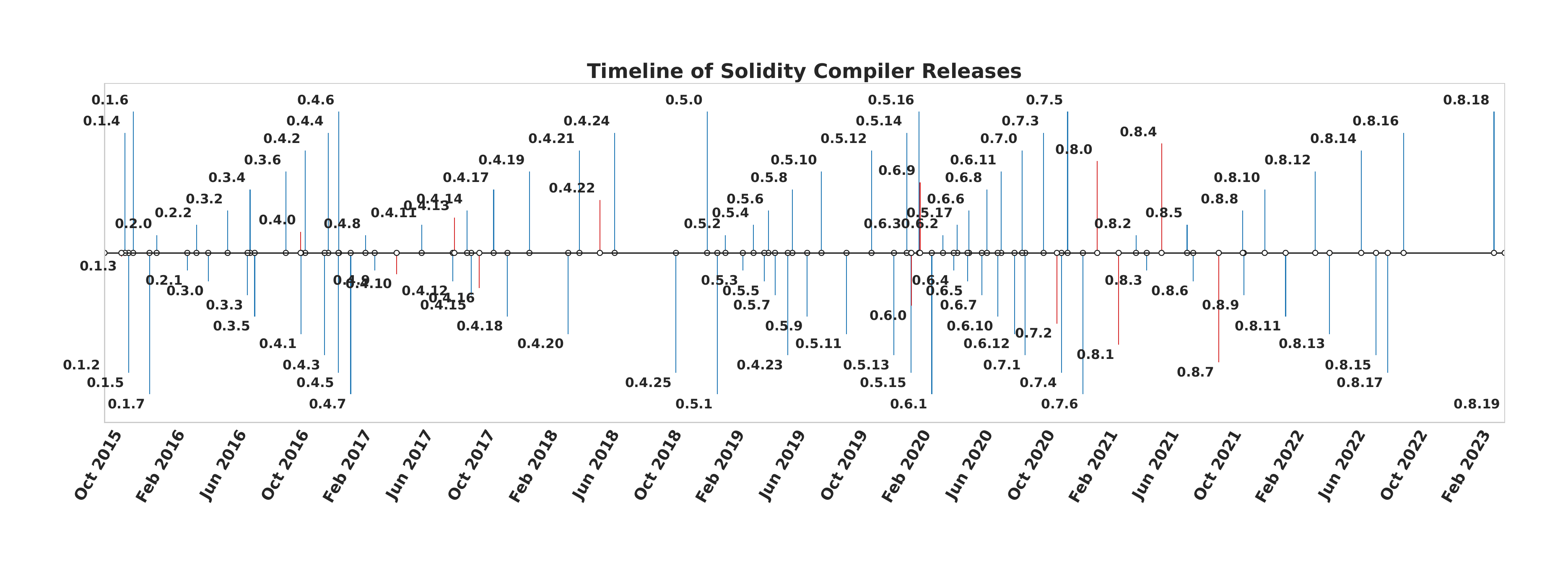}}
	\vspace{-1.5mm}
    \caption{RQ2. Timeline of the Solidity releases.
    Each line corresponds to a Solidity release.
    A red line involves a release that
    incorporates changes in error handling.}
    \label{fig:timeline}
    %\vspace{-1.5mm}
\end{figure*}

%The {\bf goal} of RQ2
%is to examine the available
%API reference documentation of the different
%versions of \code{solc} and identify changes in
%the specification usages of the Solidity {\sc EH} features.

Table \ref{table:signif_changes},
lists the most significant changes of {\sc EH} features
across the different versions of Solidity,
found by our manual analysis (see Section~\ref{sec:protocol}).
Up to \code{v0.4.10},
there are no {\sc EH} features,
except for the \code{throw} keyword.
Consider that Solidity designers deprecate \code{throw} in
\code{v0.4.13}~\cite{docsv0.4.13}.
%In the following paragraphs,
%we further discuss these changes providing
%corresponding insights.
After \code{throw}'s deprecation,
Solidity developers introduce three new
{\sc EH} features, namely,
\code{require},
\code{revert},
and \code{assert}.
\code{v0.4.22} is released
with a number of features that enable
developers to specify
messages in \code{require}
and \code{revert}.
\code{v0.4.22} introduces the
{\sc {\sc SMTChecker}},
the formal verification module of {\tt solc},
which developers could enable in
their contract via the \code{pragma} keyword.
%which could be enabled by developers
%through the smart contracts.
%This module could be enabled
%via the following statement in the beginning of
%a smart contract:
%\code{pragma experimental {\sc SMTChecker}}.
Utilising the {\sc {\sc SMTChecker}},
developers could automatically detect, for instance, 
arithmetic overflows
at compile time.
In \code{v0.6.0},
\code{try}--{\tt catch}
is introduced, and
from \code{v0.6.9} and on,
the {\sc {\sc SMTChecker}} is updated
to check array-related actions,
overflows and underflows.
Consider that until then,
such checks could be performed
via \code{require} and \code{assert}.
\code{v0.8.0} introduces
two main types of exceptions:
{\tt Panic} and {\tt Error}.
{\tt Panic} exceptions are created by the compiler in
certain situations~\cite{soldoc},
or they can be triggered by the \code{assert} function.
%On the other hand,
An {\tt Error} exception can be generated by
either {\tt require} or {\tt revert}.
Such checks are used to ensure valid conditions
that cannot be detected until execution time.
Furthermore,
from \code{v0.8.1} and on,
developers have the ability to
use {\tt Panic} and {\tt Error}
inside a {\tt catch} statement.
Then,
in \code{v0.8.4},
%the usage of:
%\code{pragma experimental {\sc SMTChecker}}
%is deprecated
%until \code{v0.8.7},
%where
the {\sc SMTChecker} runs by default.
Thus,
from \code{v0.8.7} and on-wards,
several evaluations such as
division by zero and arithmetic overflows
are performed at the compiler level.

Figure~\ref{fig:timeline}
demonstrates the Solidity releases
form October 2015 to today.
%Lines with red color,
%indicate the Solidity
%releases that include a
%change related to an {\sc EH} feature.
Consider that there is a significant number of
versions released
between February 2019 and February 2021.
This trend arises after the first appearance of
{\tt try}--{\tt catch} in Solidity,
in \code{v0.6.0}.
\begin{comment}
Then, in v0.6.4, v0.6.9, v0.8.0, and v0.8.1
Solidity developers need to
perform updates related to the newly introduced
{\sc EH} feature, {\tt try}--{\tt catch},
to ensure its stable usage.
During this period of time,
there are four versions that include
bug fixes in {\tt try}--{\tt catch} statements
(\code{v0.6.4, v0.6.9, v0.8.0, v0.8.1}),
and there are many versions ($\approx8$)
that add features or fix bugs in
{\sc SMTChecker}.
\end{comment}

The total number of \code{solc} releases
are 102 (until \code{v0.8.19}).
The median of all releases is 28.50 days between two releases,
while the average
is 27.91 days.
The total releases of the Solidity compiler that include a change in an {\sc EH} feature are 13.
The median is 112.5 days, while the average days between two releases that include a change in {\sc EH} is 179.08 days.
%The frequent releases containing
%changes related to {\sc EH}
%highlight the fact that Solidity
%is still in its early stages
%but matures quickly over time.
\vspace{-1.9mm}
\begin{rqbox}
 	{\bf Answer to RQ2:} Significant changes in the {\sc EH} features of Solidity include the deprecation of {\tt throw}, the introduction of {\tt require} and {\tt try}--{\tt catch}, as well
    as the introduction of the {\sc SMTChecker}
    that currently supports checks performed via {\tt assert}.
    On average,
    changes in {\sc EH} features happen once or
    twice per year at least.
\end{rqbox}

%% file: sections/rq3.tex
\begin{figure}
    \includegraphics[scale=0.40]{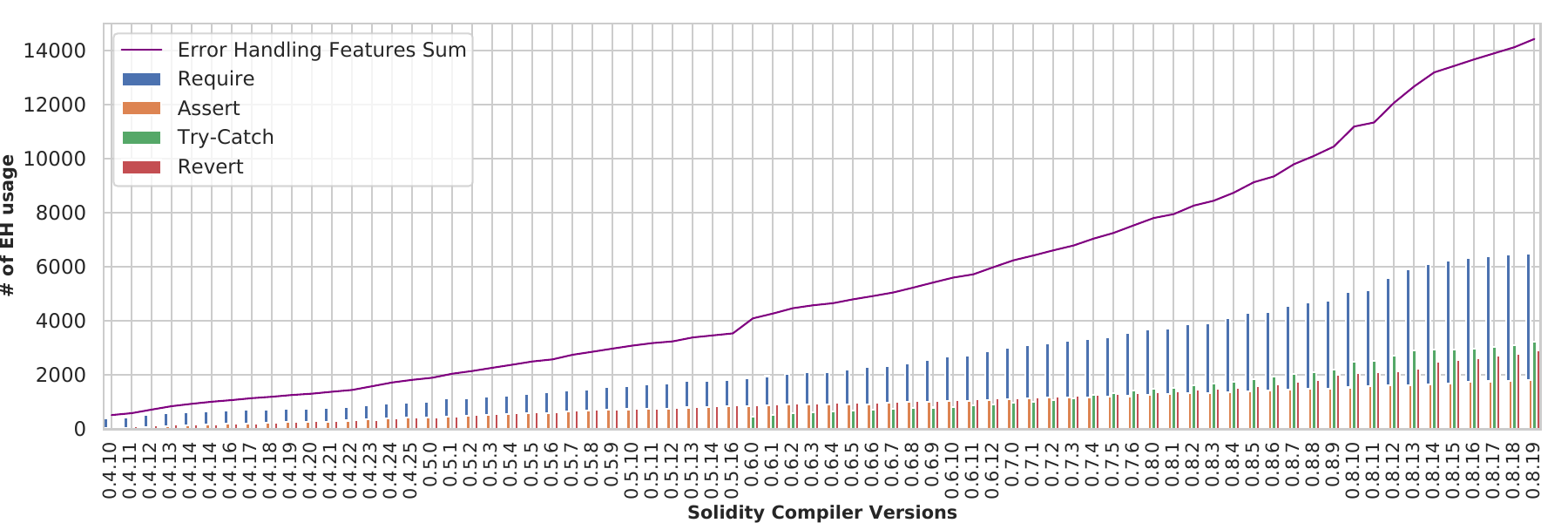}
   \vspace{-0.5mm}
    \caption{RQ3. Error-handling (EH) usage across Solidity versions.}
    \label{fig:versions}
    \vspace{-0.5mm}
\end{figure}

\begin{figure}
	\includegraphics[scale=0.40]{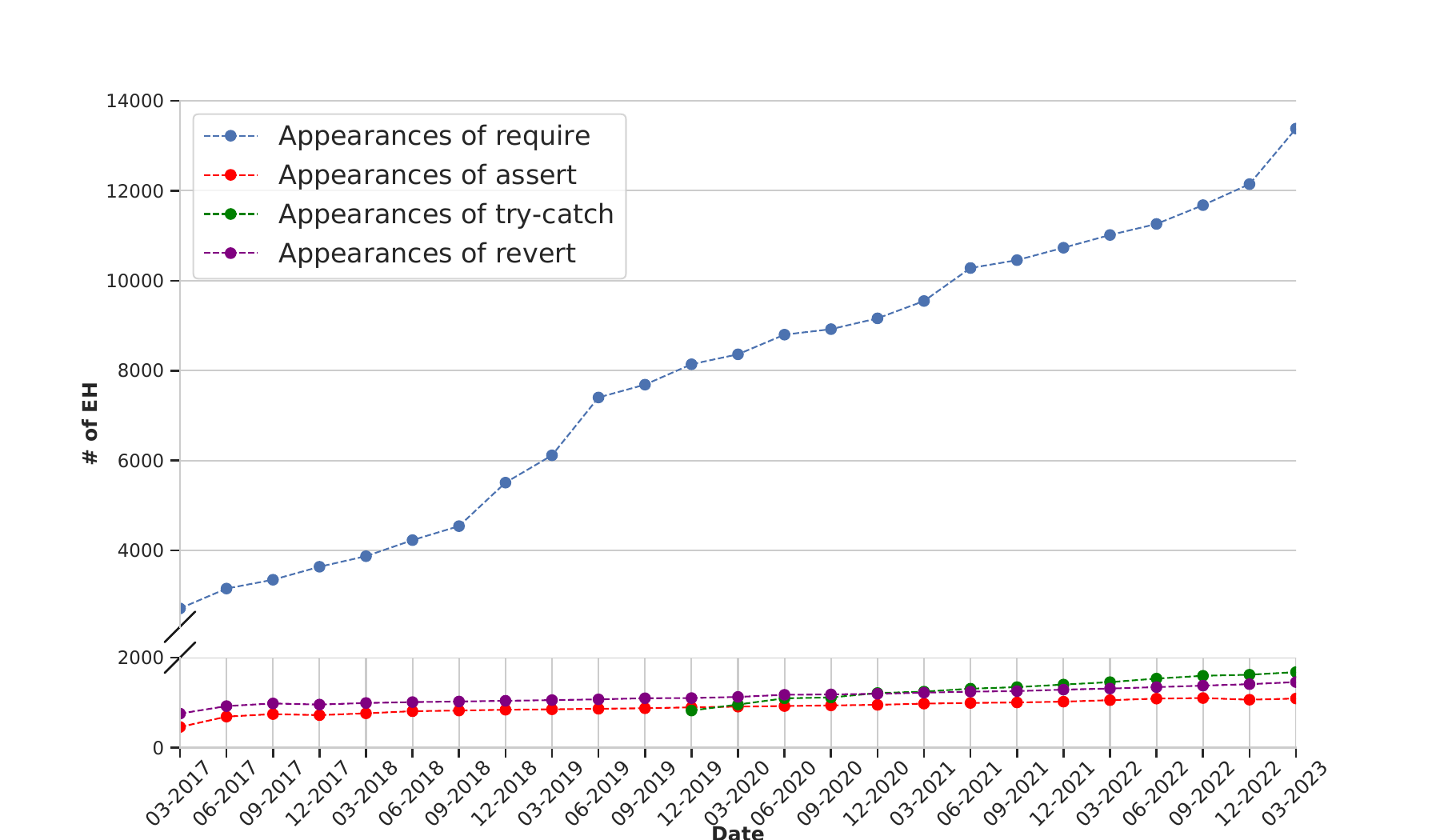}
	\vspace{-0.5mm}
    \caption{RQ3. Error-handling (EH) usage
    across years.}
    \label{fig:months}
\end{figure}

%The {\bf goal} of RQ3 is to examine how developers adopt the usage of {\sc EH} features over time and check how this evolution is related to the changes identified in RQ2.
%We include contracts that use
%\code{solc} versions from {\tt 0.4.10}
%(released in March 2017)
%and on-wards.
%This is because core {\sc EH} features such as {\tt require},
%{\tt assert} and {\tt revert} were introduced in {\tt v0.4.10}.

% \vspace{-8.0mm}
Figure \ref{fig:versions}
illustrates how many times
each {\sc EH} feature is used in our dataset,
across the different Solidity versions.
Also,
it presents how many times all features are used per version
(i.e., {\sc EH} features' sum).
If a smart contract with {\sc EH} features can be compiled with {\tt solc}
$\ge$ \code{v0.4.10},
then we count each detected feature
in this specific smart contract,
for the whole range of these versions.

Our findings indicate that \code{require}
is the most frequently used {\sc EH} feature
across all Solidity versions.
In the first version, where
{\sc EH} features appear
(\code{v0.4.10}),
corresponding smart contracts include
410 instances of {\tt require},
76 instances of {\tt revert},
and 31 instances of {\tt assert}.
In the last version considered,
i.e.,
\code{v0.8.19},
\code{require}
appears 6,484 times,
while \code{assert} appears 1,813 times,
\code{revert} 2,901 times, and
\code{try}--{\tt catch}
3,231 times.
Overall,
developers seem to use all features more
as time goes by.
An interesting observation involves
\code{try}--{\tt catch}.
The feature is introduced in \code{v0.6.0}
(see Table~\ref{table:signif_changes}),
and its usage is gradually increasing until
\code{v0.7.4}
(appearing 1,275 times in this version)
and eventually takes the lead
from all the other
{\sc EH} features, except for \code{require}.

Figure \ref{fig:months},
illustrates the frequencies of
{\sc EH} feature usages
across time,
per quarter.
%Consider also Table \ref{table:signif_changes}
%for the {\sc EH} changes across versions.
%starting from the first version considered in
%our study, \code{v0.4.10}
Overall,
\code{require}
always appears at least 1,000 times more than
\code{revert} and \code{assert}.
A sharp increase of the
{\tt require} usage
(2,500 times)
occurs from September 2018
to June 2019.
The increase
coincides with the release of \code{v0.4.22},
where Solidity provides developers with the ability
to specify error messages in \code{require}
%code fragments 
(see Table \ref{table:signif_changes}).
Furthermore,
\code{revert} is more utilised
than \code{assert} throughout the timeline.
After the introduction of \code{try}--\code{catch}
in \code{v0.6.0},
in 2019
(see Table \ref{table:signif_changes}),
\code{try}--\code{catch}
counterbalances
the other two {\sc EH} features,
i.e.,
\code{revert} and \code{assert}.
This happens from September 2019 to March 2020.
In June 2020,
the usage of \code{try}--\code{catch}
surpasses the usage of
\code{revert} and \code{assert}, respectively.
The lead becomes more obvious
in January 2021,
when Solidity \code{v0.8.1} is released.
A reason behind this
may involve a
newly introduced
%This may happens due to a newly introduced
\code{try}--\code{catch} property
that provides developers with the ability
to specify
the reason of a failure
using \code{Panic} and \code{Error}
(see Table \ref{table:signif_changes}).
\setlength{\textfloatsep}{10pt} % Adjust the value as needed
\begin{table}[t]
\centering
\caption{RQ3. Error-handling (EH) usage growth rates.}
\vspace{-1mm}
\resizebox{\columnwidth}{!}{
\begin{tabular}{ r | r | r | r | r | r }
    \toprule
   {\bf EH feature} & {\bf Min}  & {\bf Max} & {\bf Average} & {\bf Median} & {\bf Standard Deviation}\\
   \midrule
   % \rowcolor{gray!25}
    \code{require}&0.163&3.938&1.894&2.085&1.187\\
    \code{try-catch}&0.162&1.037&0.333&0.162&0.378\\
    % \rowcolor{gray!25}
    \code{revert}&0.220&0.927&0.513&0.490&0.224\\
    \code{assert}& 0.504&1.406&0.961&0.989&0.329\\
    \bottomrule
\end{tabular}
}
\label{table:ehfeaturesstatistics}
\end{table}

Table~\ref{table:ehfeaturesstatistics}
describes that on average
the use of {\tt require} has the highest increase
over time.
Then,
{\tt assert} and {\tt revert} follow.
The most recent {\sc EH} feature introduced in Solidity,
i.e., {\tt try}--{\tt catch},
presents the lowest, but increasing, growth rate over time.
\begin{rqbox}
    {\bf Answer to RQ3:} \code{require}
    has the highest increase across versions and years.
    From the second half of 2018,
    {\tt require} has a sharp increase,
    which coincides
    with the release of {\tt v0.4.22},
    where developers can specify
    error messages 
    in {\tt require} and {\tt revert}.
    With the introduction of \code{try}--\code{catch}, in {\tt v0.6.0},
    instantly, its usages
    become equal to those of \code{revert} and \code{assert},
    surpassing them since June 2021.
\end{rqbox}

%% file: sections/rq4.tex
%The {\bf goal} of RQ4 is
%to identify and
%measure Solidity {\sc EH} misuses.
%Again here,
%we consider contracts
%working with \code{solc} \code{v0.4.10}
%and on-wards for the same reasons
%discussed in Section~\ref{sec:usage_evol}.

Table~\ref{table:misuses}
categorises {\sc EH} misuses
providing corresponding descriptions,
the total cases where {\sc EH}
should have been used,
and the number of misuses
(i.e., missing {\sc EH})
for each category.
In the following,
we discuss in detail the misuse categories
we identify.

{\bf External Call ({\sc ECall}).}
When a smart contract performs
a transaction,
it usually makes calls to
external contracts.
According to the documentation,
such calls must
be verified by developers
using {\sc EH} features
such as \code{require} or
{\tt try}--{\tt catch},
otherwise,
potential problems may occur.
For instance,
if developers ignore checking
whether an external call
(e.g.,
{\tt someAddress.call();})
can return a zero {\tt address},
then the call may return results
that can modify the state
of the caller contract in an
unexpected manner.
Examining our dataset,
we found
110,655 external calls.
For a significant number of external calls ($63.63\%$),
developers use neither \code{require} nor {\tt try}--{\tt catch}.
Therefore, those situations represent potential software failures
that happen due to an unchecked call return value~\cite{SWC-104}.
Furthermore, consider that if an external call fails, it should be isolated (i.e., via {\sc EH}), otherwise it can cause a DoS (Denial of Service) vulnerability~\cite{SWC-113}.

\begin{table}

\caption{RQ4. Categories of error-handling (EH) misuses,
corresponding descriptions, total cases where EH should have been used, and number of misuses for each category.
}
\vspace{-1mm}
\resizebox{1.03\columnwidth}{!}{
\begin{tabular}{ p{1.9cm} |  p{2.1cm} |  p{2.2cm} |  p{4.0cm} | p{1.0cm} | p{2.0cm} | p{1.2cm}}
    \toprule
   {\bf Missing EH Category (EH Misuse Category)} & {\bf Category Description}  & {\bf Representative Misuse (Contract in Etherscan)} & {\bf Misuse Description} & {\bf Total Cases (\#)} & {\bf Missing EH (\%)} & {\bf EH Required}\\
   \midrule
   \rowcolor{gray!25}
    {External Call ({\sc ECall})}&Missing EH feature to validate a call to an external contract& {\tt IRoutedSwapper} & Missing \code{require} or {\tt try}--{\tt catch} for external call \code{swapper.swapExactInput())}&110,655& $70.417 (63.63\%)$&\code{require} or {\tt try}--{\tt catch}\\
    \midrule
    {Function Argument of \code{address} type ({\sc FAA})}& Missing \code{require} to check
    if an input of an \code{address} type is valid.&{\tt StreetDawgs}&Missing \code{require} check in the code fragment \code{address from = address(uint(pOwner(Id))}&31,018& $27,741 (89.43\%)$&\multirow{3}{*}{\code{require}}\\
    \midrule
    \rowcolor{gray!25}
    {External Contract ({\sc ECon})}&Missing \code{try} to test if the external contract is created correctly &{\tt CapsuleFactory}&A new contract \code{Capsule capsuleCollection = new Capsule(name, symbol, tokenURIOwner, isCollection)} is created without using \code{try} to test the correct creation of the contract &1,162& $934 (80.37\%)$&\multirow{4}{*}{\code{try}--{\tt catch}}\\
    \midrule
    {Array Allocation ({\sc AA})}&Missing \code{assert} to test if the array allocated has a valid length &{\tt VaultsRegistry}&Missing \code{assert} to test if \code{allVaults.length}
    in \code{address[] memory vaultsArray = new address[](allVaults.length);} is a valid number &602& $401 (66.67\%)$&
    \multirow{4}{*}{\code{assert}}\\
    \rowcolor{gray!25}
    {Pop from Array ({\sc PA})}&Missing \code{assert} to test if \code{.pop()} is valid & {\tt GenjiMonotagari}& The function \code{remove(Map storage map, address key)} that deletes the last key from \code{map.keys} does not have an assertion.&181& $164 (90.62\%)$&\multirow{3}{*}{{\code{assert}}}\\
    {Division by Zero ({\sc DZ})}&Missing \code{assert} to check if the division is valid &{\tt ZoomerRethPool}&In function \code{getRedeemValue(uint256 totalBurn)}
    there is not check if the \code{(balance / (5000 - totalBurn)} is  \code{>0}&94& $75 (79.76\%)$&\multirow{3}{*}{{\code{assert}}}\\
    \rowcolor{gray!25}
    {{\tt enum} Type Conversion ({\sc ETC})}&Missing \code{assert} to validate if the conversion to the \code{enum} type is done correctly &{\tt Hakiro}&In this contract, there is an {\tt enum} type named \code{Step}. In function \code{setStep(uint step)}, there is a missing assertion to check if \code{sellingStep = Step(step);} is done correctly &67& $60 (89.55\%)$&\multirow{4}{*}{{\code{assert}}}\\
    \bottomrule
\end{tabular}
}
\label{table:misuses}
\end{table}

{\bf Function Argument of
{\tt address} type ({\sc FAA}).}
To perform transactions,
contracts use the {\tt address} type
to define the address of the sender and
the recipient.
According to the documentation,
the {\tt require} {\sc EH} feature
should be used to guarantee
that the values of the {\tt address} type
are valid.
For example,
if the sender sends an amount to
a receiver that has zero {\tt address},
this amount will be sent to an account that does not exist
and a loss of funds will occur.
To ensure that the value of {\tt address} is non-zero,
developers need to use {\tt require},
i.e., {\tt require(recipient != address(0),"...");}
In our dataset,
we detect 31,018 coding situations where the
function argument of {\tt address} type
should be verified by developers.
For 89.43\% of those situations,
developers do not
check if the address is
not zero using \code{require}.
Such cases can lead
to attacks such as the
the Nomad bridge attack~\cite{nomad}
described in Section~\ref{sec:background}.

{\bf External Contract ({\sc ECon}).}
In Solidity,
developers have the ability to
create new smart contracts
within other smart contracts, 
using the \code{new} keyword.
Without using the {\tt try}--{\tt catch} {\sc EH} feature,
the creation of a contract within another contract
can lead to software failures.
Consider a smart contract A that uses a function
to create a new smart contract B.
If the deployment process is not
done correctly and encounters an error,
B may not be fully initialised,
or might be left in an incomplete state,
producing an exception.
If {\tt try}--{\tt catch} is not used,
the exception remains ``uncaught'',
leading to an unexpected behaviour within the smart contract A.
Overall,
we identify 1,162 coding situations
of external contract in our dataset.
From those situations,
80.37\% do not involve \code{try}--{\tt catch},
and can potentially cause software failures.
Consider that there are also related situations
that a vulnerability can happen
due to an
incorrect constructor name~\cite{SWC-118}.

{\bf Array Allocation ({\sc AA})}.
In Solidity, one
can deploy a smart contract which allocates a memory array of user-supplied length.
According to the documentation,
if the length of a new array
is not verified using {\tt assert},
and the array's length is invalid,
software failures may occur,
including memory corruption~\cite{memcor}.
From the 602 situations where
array allocation takes place in our dataset,
66.67\% do not involve {\tt assert}.

{\bf Pop from Array ({\sc PA}).}
In Solidity,
developers can delete elements from arrays using the {\tt pop} function.
However, if developers ignore
using {\tt assert}
to test if the array is empty,
the contract's execution will be stopped
unexpectedly.
Analysing our dataset,
we found that
in 181 coding situations
where {\tt pop} was used
to delete elements from arrays,
90.62\% do not use \code{assert},
as suggested by the documentation
(for Solidity  $<$ \code{v0.8.7}).

{\bf Division by Zero ({\sc DZ}).}
For division by zero,
developers should use
\code{assert} according to the
documentation
(again for versions $<$ \code{v0.8.7}).
A division by zero
can end the execution of the smart contract
unexpectedly.
In our dataset,
we found 94 cases where
division by zero may occur.
However,
79.76\% of them do not use {\tt assert}.
Such situations
%of bad code,
%which include division by zero,
%or other incorrect calculations
%(e.g.,
%integer overflow/underflow~\cite{SWC-101}),
can be considered as potential
exploits~\cite{division}.

{\bf {\tt enum} Type Conversion ({\sc ETC}).}
According to the Solidity's documentation,
developers should
use \code{assert} when they try to
convert a value that is too big or negative into an
{\tt enum} type.
Consider a function
that has a \code{uint} value
and tries to convert this value
into an \code{enum} value.
If the conversion is unsuccessful,
there is a possibility that
this value
will have a different type.
To guarantee that the conversion
will be successful,
developers should use \code{assert}.
In our dataset,
from the 67 coding situations that
include \code{enum} type conversions,
89.55\% do not check the validity of the conversion.

Overall,
the misuse category
that developers mostly ignore error handling is
``pop from array'' ({\sc PA}, 90.62\%).
This issue has been addressed though
after the introduction of the {\sc SMTChecker},
in Solidity \code{v0.6.9},
where {\tt solc}
automatically checks for invalid {\tt pop} calls.
Developers still do not use error handling for other categories such as
``\code{enum} type conversion''
(89.55\%) and ``function argument of \code{address}
type'' ({\sc ETC}, 89.43\%).
The category for which developers mostly
use error handling is the
``external calls'' ({\sc ECall}),
where we find the lowest number of misuses
(63.63\%).

\begin{rqbox}
	{\bf Answer to RQ4:}
    In total, the category with the highest number of total cases that require
    error handling is the ``external call'' (110,655).
    However, for the ``external calls'' developers
    use error handling (63.63\%) more than the other categories.
    The categories that developers mostly ignore error handling include
    ``pop from array'' (90.62\%), ``{\tt enum} type conversion'' (89.55\%), and
    ``function argument of {\tt address} type'' (89.43\%).
\end{rqbox}

%% file: sections/rq5.tex
\begin{figure}
\centering
    \includegraphics[scale=0.30]{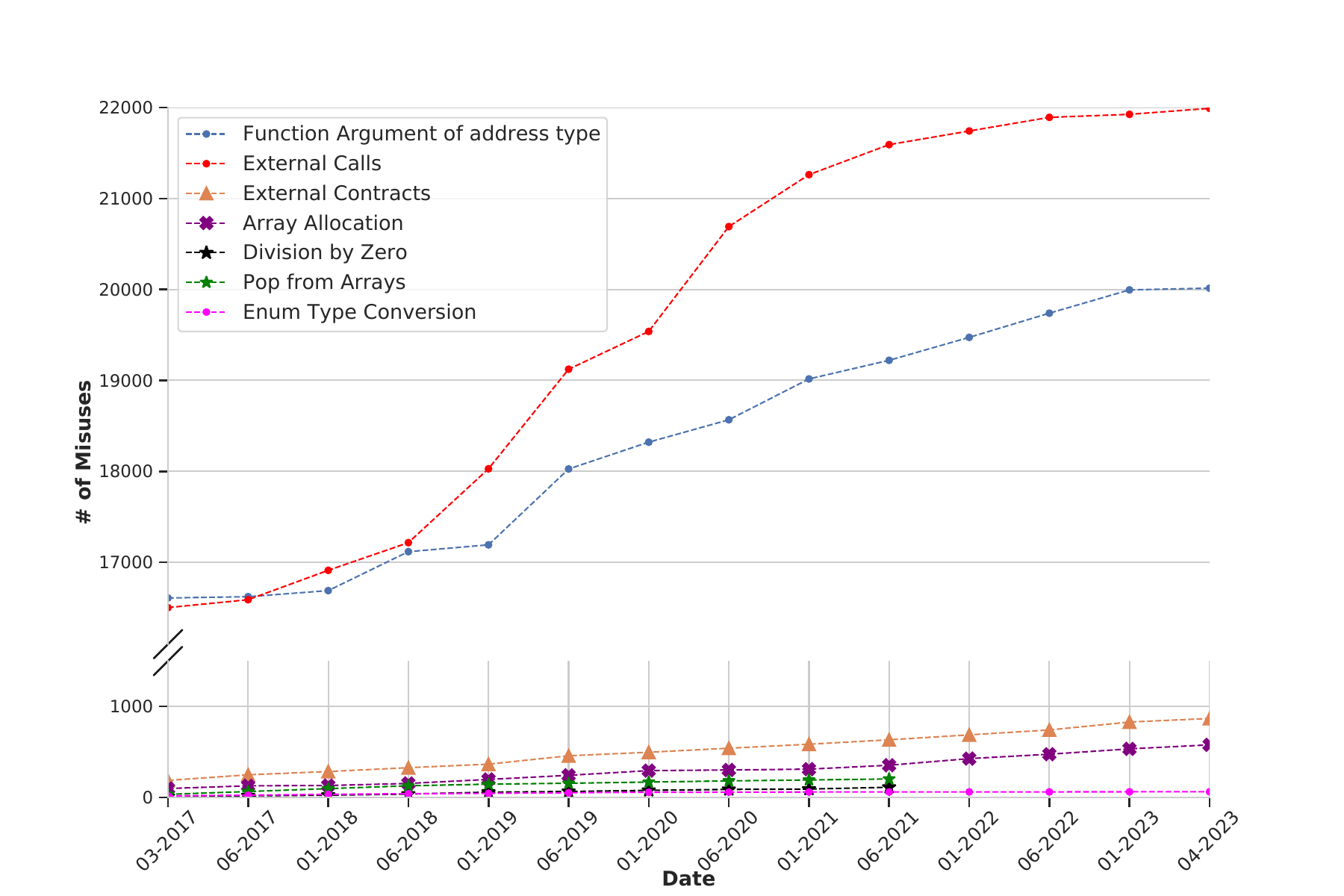}
    \vspace*{-1.5mm}
    \caption{RQ5. Error-handling (EH) misuses per quarter.}
    \label{fig:misugesovertime}
\end{figure}
%The {\bf goal} of RQ5 is to examine how {\sc EH} misuses
%evolve over time in the Solidity context.
%As before,
%we consider contracts
%working with Solidity \code{v0.4.10}
%and on-wards
%(see Section~\ref{sec:usage_evol}).

Figure \ref{fig:misugesovertime},
illustrates how the misuses identified in our dataset
evolve over time,
per quarter.
In the first quarter of 2017,
the number of ``function argument of \code{address} type''
({\sc FAA}) is 926 times higher than ``external call''
({\sc ECall}).
In June 2017,
{\sc ECall}'s frequency becomes
higher than {\sc FAA}'s for the first time.
From that point and on,
both have a steady increase over time
with {\sc ECall} leading the way.
At the end (04-2023),
the number of {\sc ECall}
is 31,277 higher than the
{\sc FAA}.
The third most common misuse is
``external contract''
({\sc ECon}),
being steadily $\approx127\%$
times higher than ``array allocation''
({\sc AA}).
Through time,
{\sc AA} is
on average $\approx76\%$ higher than 
``division by zero''
({\sc DZ}),
and $\approx85\%$ higher than 
``{\tt enum} type conversion''
({\sc ETC}).
Note that {\tt assert}-related misuses
({\sc DZ} and ``pop from array'' -- {\sc PA})
frequencies are zeroed after June 2021.
This happens due to the incorporation of the
{\sc SMTChecker} that runs by default
in {\tt solc} \code{v0.8.7},
and was released in June 2021.
Recall that the {\sc SMTChecker}
automatically performs checks for {\sc DZ} and {\sc PA},
nullifying the use of {\tt assert} for the corresponding coding situations.

% \vspace{-2mm}

Table~\ref{table:misusesstatistics}
presents descriptive statistics
regarding the growth rates of {\sc EH} misuses per quarter.
{\sc FAA} and {\sc ECall} have the lowest median.
This may seem slightly contradictory to the findings related to the corresponding misuses frequencies.
However,
the low growth rates involve the
rise of using {\tt require}
over time to handle external calls and function
arguments (see also Sec.~\ref{sec:usage_evol}).
{\sc ECon} has the next lowest median,
meaning that developers relatively use
{\tt try}--{\tt catch} to examine the
creation of an external contract.
{\sc DZ} and {\sc PA} have the highest median.
This is because developers do not use {\tt assert}
to assess either if {\tt pop} is valid or
if a division by zero may take place.
Note that in the case of {\sc DZ} and {\sc PA},
we examine the period of time prior to
the inclusion of the {\sc SMTChecker}.
The next categories that have the
highest median include
{\tt ETC} and {\tt AA}.
This indicates that developers should consider
using {\tt assert} more often to check
array lengths and {\tt enum} type conversions.
Recall that {\sc SMTChecker} have yet to automatically
check the corresponding coding situations.

\begin{rqbox}
	{\bf Answer to RQ5:}
Over time, the highest number of 
misuses involves the absence of {\sc EH} for (1)
validating calls to external contracts,
(2) checking if the input of an
{\tt address} type is reliable.
However, the corresponding growth rates are low,
which indicates the increasing use
of the {\sc EH} features that
can handle such cases.
The impact of the introduction of the
{\sc SMTChecker} into the Solidity compiler is prominent here,
with corresponding misuses' frequencies
%(e.g., missing division-by-zero checks)
dropping ever since its introduction.
\end{rqbox}

\begin{table}[t]
\centering
\caption{RQ5. Error-handling (EH) misuse growth rates.}
\vspace{-1mm}
\resizebox{0.99\columnwidth}{!}{
\begin{tabular}{ l | r | r | r | r | r}
    \toprule
   {\bf EH Misuse Category} & {\bf Min}  & {\bf Max} & {\bf Mean} & {\bf Median} & {\bf St. Dev.}\\
   \midrule
   % \rowcolor{gray!25}
    {External Call ({\sc ECall})}&0.043&0.332&0.190&0.218&0.133\\
    {Function Argument of \code{address} type ({\sc FAA})}& 0.004&0.205&0.103&0.110&0.110\\
    % \rowcolor{gray!25}
    {External Contract ({\sc ECon})}&0.331&3.641&1.777&1.775&1.164\\
    {Array Allocation ({\sc AA})}&0.306&4.897&2.081&2.045&1.590\\
    % \rowcolor{gray!25}
    {Pop from Array ({\sc PA})}&0.911&4.970&3.588&4.161&1.646\\
    {Division by Zero ({\sc DZ})}&0.801&10.201&6.364&7.299&3.832\\
    % \rowcolor{gray!25}
    {{\tt enum} Type Conversion ({\sc ETC})}&0.470&2.647&1.920&2.382&0.881\\
    \bottomrule
\end{tabular}
}
%\%vspace{1mm}
\label{table:misusesstatistics}
\end{table}

%% file: sections/discussion.tex
\begin{comment}
In the following paragraphs,
we summarise the key take-aways of our study
and make suggestions for improving
Solidity and its documentation.
However,
we acknowledge that our suggestions should
be further validated in future work,
possibly by involving researchers and practitioners.
%and for Solidity
%designers and smart contracts' developers.
\end{comment}

%\begin{itemize}
%\item
{\bf Amplifying EH usage in Solidity}.
Our results indicate that overall,
the usage of most {\sc EH} features
over time is limited
(except for {\tt require}
that follows an upward trend).
Thus, we call for actions
that can assist developers
to engage in using those features.
Such actions may include
the improvement of the Solidity documentation,
the development of ad-hoc tools that can automatically suggest
{\sc {\sc EH}} usages,
and the addition of extra checks into the {\sc SMTChecker}.

%\item
{\bf Stabilising Solidity's volatile nature}.
Our work highlights
trends in the evolution of error handling in Solidity.
Solidity designers change quite often {\sc EH} features
making Solidity volatile.
Then, maybe it is difficult for developers
to follow those changes, and ignore {\sc EH} features.
Thus, similar future studies to ours,
as well as human surveys, could help Solidity designers to
understand potential needs and, consequently, stabilise their framework.

%\item
{\bf Enhancing Solidity documentation}.
The {\sc EH} misuses we discussed in Section~\ref{sec:rq3}
reveal that developers ignore or may find difficult
to understand the specification usage of {\sc EH} features in the documentation.
Therefore, there is a need for improving the documentation
in terms of explainability,
so that developers can understand the usage of {\sc EH}
features and remain engaged in {\sc EH} usage
(e.g.,
how to safely handle elements involved in transactions
among smart contracts).
Solidity designers could
also add to the documentation
representative examples of real-world {\sc EH} misuses and their impact in terms of reliability and security.
We include such examples in Section~\ref{sec:rq3}.
{\bf Introducing data-flow analysis}.
Incorporating data-flow analysis into \code{solc}
could also help developers write more secure smart contracts.
%Given the high frequency of external calls found in smart contracts,
%and the need for protecting such calls using \code{require},
%data-flow analysis can be incorporated into \code{solc}.
By employing data-flow analysis,
developers will be able to
to detect inputs coming from external calls,
track their flow within the contract,
and check if they reach any sensitive ``sinks''.
Based on this information,
they will be able to automatically identify where
the use of \code{require} is essential.
Furthermore,
data-flow analysis can be performed by ad-hoc tools
as done in other programming languages such as Java~\cite{ARF14}.
%\end{itemize}

%% file: sections/threats.tex
\label{sec:threats}

Our work may suffer from both internal validity threats (related to the implementation of {\sc SolBench}) and external validity threats (related to the generalisation of our findings).
In the following,
we discuss these threats and our
attempts to mitigate them.

{\bf Internal validity.}
A potential internal threat involves
the false positives
of {\sc SolBench}.
To tackle this issue,
two authors manually checked for false positives in the findings (e.g.,
whether a misuse identified by {\sc SolBench}
is not an actual misuse).
Specifically, for each misuse category,
we randomly selected ten corresponding cases from our findings.
Then, two authors examined each misuse to
check for false positives.
No false positives were identified.
However, throughout the process,
we made an interesting observation.
In some occasions,
developers attempted to reproduce
{\sc EH} functionalities
(e.g., check for zero {\tt address})
by using {\tt if} statements.
Note that such cases cannot be classified
as {\sc EH} misuses.
As in all qualitative studies,
our manual analysis of the 
Solidity documentation
discussed in Section~\ref{sec:protocol},
may suffer from human errors.
To mitigate this issue
the authors cross-checked their findings
following the process explained
at the end of Section~\ref{sec:protocol}.
%To address this issue,
%two authors
%(other than the two
%that performed the initial analysis)
%cross-checked the results.
%and we made our results
%publicly available [6, 8].
Finally,
we did not examine potential
{\tt revert} misuses.
Recall that this {\sc EH} feature
is mainly used to revert a transaction
if a condition is not met.
To do so, developers have to include it
in {\tt if}--{\tt else} conditionals that are tailored to the smart contract's logic.
Given this characteristic,
it is not straightforward for one to identify a misuse involving {\tt revert}.

{\bf External validity.}
We derived our results based on the
dataset we curated (see Section~\ref{sec:dataset}).
We plan to run {\sc SolBench}
using other existing benchmarks~\cite{RYM21,PL21,DFA20}
to strengthen its generalisability.
Consider though that we already incorporated
the benchmark of Chaliasos et al.~\cite{CGL22} in our dataset.
Finally, we examined all \code{solc} versions
(\code{v0.1.2}--\code{v0.8.19}) that have available documentation
until the date of our study.
Therefore, for newer Solidity versions our results may differ.

% \noindent
% {\bf Availability.}
% One can reproduce the results
% of our study by accessing
% \dataset{} online. \footnote{\url{https://github.com/Solidity-ErrorHandling-Anonymous/solbench}}
% We plan to make a corresponding eponymous repository
% publicly available upon acceptance.
% \todo{if we have space let's move this in the last section "DATA AVAILABILITY" after the conclusions section}

\begin{comment}
Searching for {\it misuses}
with our {\it heuristic rules},
we try to detect for {\bf false positives}:
Process - Findings:
\begin{itemize}
    \item 2 validators
    \item run the {\it heuristic rules} for all the contracts in the dataset
    \item examine manually 10 cases of each {\it misuse} category
    \item in all the cases detected from the {\it heuristic rules}
    there are only correct {\it misuse} cases (no false positives).
    \item * manually observe usage cases (found with {\it heuristic rules})
    that instead of {\sc EH} developers use \code{if} statements.
\end{itemize}
\end{comment}

%% file: sections/rw.tex
% Empirical studies in solidity and smart contracts
% Tools that identify misuses, bugs in Solidity
% EH in other PLs
%The following paragraphs discuss representative empirical studies
%related to our study
%and show a comparison with our work.
{\bf Studies on smart contracts.}
Several studies refer to
the automated detection and fixing of bugs
in smart contracts.
\citet{DAC20} propose {\it Smartbugs},
an extendable benchmark having
more than 40K smart contracts.
The benchmark enables the integration
and evaluation of different security tools
that can analyse smart contracts.
\citet{CCZGGML23} introduce an approach
that utilizes the Smartbugs benchmark and questionnaires
to further examine the effectiveness and usage of
security tools against
real-world vulnerabilities.
\citet{RYM21} propose a four-step evaluation method
for minimising bias in the assessment of static tools.
The dataset used for their experiments contains
46,186 smart contracts.

Other studies,
focus on particular
features of the Solidity programming language.
\citet{CGL22} conduct an empirical study
on 50M smart contracts
%extracted from Etherscan
to explore the use of inline assembly.
Similarly,
\citet{LSZ23}
examine the use of inline assembly in more
than 7.6M open-source Ethereum smart contracts.
\citet{LWZ21} conduct a large-scale study,
on 3,866 smart contracts,
to identify the use of transaction-reverting statements.
Consider that such statements involve {\sc EH} features
such as {\tt require} and {\tt revert}.
Our study is broader though, focusing on different {\sc EH} aspects,
examining {\sc EH} features' frequency and evolution.

Several studies investigate
bugs and vulnerabilities in relation to smart contracts.
\citet{ABC17} perform an extensive survey of smart contract attacks.
\citet{PL21}
survey more than 23K smart contracts reported
as vulnerable in different academic papers,
indicating a high number of false alarms in existing static tools.
\citet{ZZXL23} investigate (1) 462
defects reported in corresponding
audits done by the
well-known {\it Code4rena} contest~\cite{codearena},
and (2) 54 exploits,
to study if state-of-the-art static
tools could identify them.
\citet{YXY23} perform an empirical study on
historical bug fixes from 46 real-world Solidity smart contracts.
\citet{CXL22} identify 20 kinds of contract defects
into potential security, availability,
performance,
maintainability and reusability problems
in 587 real-world smart contracts.
Furthermore, \citet{OHJ20} conduct an
exploratory study on smart contracts to
identify software quality characteristics
such as the rate of comments and code complexity.
\citet{HR20} attempt to understand the
security level of smart contracts in the wild,
by empirically studying 55,046 real-world Solidity smart contracts.
Finally, \citet{ZLW23}
perform a study
and propose an approach to automatically detect inconsistencies
between documentation 
and corresponding code in Solidity smart-contract libraries.
%Their approach reveal high-priority
%API documentation errors in popular smart-contract libraries,
%such as parameter mismatches,
%missing requirements, and outdated descriptions.

\noindent
{\bf Studies on error handling.}
%Regarding error handling in Solidity,
%there are only a few studies available.
%Specifically,
\citet{VRH20}
use a dataset of 26,799 smart contracts
to examine the usage of Solidity functions
such as
{\tt call}, {\tt send}, and {\tt transfer}.
%which can be used to exchange cryptocurrency among contracts.
To evaluate whether those functions are used
in a secure manner,
the study also investigates the use of three Solidity guards,
i.e.,
{\tt assert}, {\tt require}, and {\tt revert}.
%that can be used to stop a transaction,
%and prevent possible exploits related
%to {\tt call}, {\tt send}, and {\tt transfer}.
As in our study,
\citet{VRH20} also find that
the developers of smart contracts mostly use {\tt require}.
Additionally,~\citet{WCZHZWJ21}
analyse 172,645 real-world smart contracts
to examine features of the Solidity programming language
related to control flow,
object-oriented programming,
data structures and error handling.
%In the control-flow features,
%the study considers EH features
%such as {\tt assert}, {\tt require}, {\tt revert},
%and {\tt try}--{\tt catch}.
This study highlights that the developers of smart contracts
extensively use {\tt assert}, {\tt require}, {\tt revert}.
In addition,
the study indicates that {\tt try}--{\tt catch} is scarcely employed.
Contrary to the aforementioned studies,
we conduct a thorough examination of
the evolution of each Solidity {\sc EH} feature and its usage
on a dataset of 283K unique open-source smart contracts.
Further,
we identify misuses of EH features,
and make relevant suggestions for their correct usage.
%We categorize the usages and misuses of 
%{\sc EH} features by cross-referencing Solidity's official documentation 
%and propose heuristic rules for their identification. Moreover,
%we scrutinize the evolution of both usages and misuses over time,
%drawing insights from a substantial dataset comprising 283,000 real-world smart contracts.

There are several studies that examine the
EH mechanisms of other
programming languages,
including Java~\cite{KGM16,NHT16,DNR13,SGH10,RM00,Kin06,Mar13,WN08,ECS15},
C++~\cite{BG19,BC15,ZLH23}, Python~\cite{PZ21},
Ada~\cite{RSB01}, Swift~\cite{CPCSA18}, and Rust~\cite{ECSL20}.
Most studies show that developers neglect
the usage of error handling.
In our study, we also find that developers
use Solidity EH features rarely.

%% file: sections/conclusions.tex
Our analysis of 283K unique smart contracts
reveals that overall
the usage of most {\sc EH} features over time is limited,
although there is an upward trend in
the usage of {\tt require}.
The popularity of {\tt require} indicates
that programming language designers
could consider the development of
{\sc EH} features that are more tailored
to the purposes of each language.
Furthermore,
our analysis on the
different versions of the Solidity documentation,
as well as the analysis of the
{\sc EH} misuses found in real-word smart contracts,
indicates that Solidity changes frequently
and has a volatile nature.
Frequent release cycles may confuse
developers of smart contracts,
making it hard for them to remain up-to-date with the
documentation of the language,
and,
thus correctly use its {\sc EH} features.
A future direction could involve human studies
that will highlight developer needs regarding
the explainability of the documentation.
Additionally,
smart contracts are particularly
exposed to {\sc EH} misuses, caused, for instance,
from unchecked external calls.
Therefore, the use of {\sc EH} features
should be further strengthened
via a more informative documentation
(e.g.,
with details about the impact of {\sc EH} misuses)
to engage developers in correct {\sc EH} usage.
We hope that our
findings can provide
valuable insights and guidance for practitioners
and researchers working in the field.